\documentclass[longnamesfirst,apjl]{emulateapj}

\usepackage{graphics}
\tighten

\usepackage{amssymb,amsmath,amsfonts}
\usepackage{color}

\newcommand{\beq}{\begin{equation}}
\newcommand{\eeq}{\end{equation}}
\newcommand{\bea}{\begin{eqnarray}}
\newcommand{\eea}{\end{eqnarray}}
\newcommand{\ba}{\begin{array}}
\newcommand{\ea}{\end{array}}

\newcommand{\lpr}{\ell^{\prime}}
\newcommand{\mpr}{m^{\prime}}

\newlength{\sizeonefig}
\newlength{\sizetwofig}
\usepackage{multirow}

\begin{document}

\title{Missing power vs low-$\ell$ Alignments in the Cosmic Microwave Background:\\ No Correlation in the Standard Cosmological Model}

\author{Devdeep Sarkar$^1$, Dragan Huterer$^1$,  
Craig Copi$^2$, Glenn Starkman$^2$, Dominik Schwarz$^3$}  

\affiliation{
$^1$ Department of Physics, University of Michigan, 450 Church St, Ann Arbor, MI 48109, USA\\
$^2$ CERCA/Department of Physics, Case Western Reserve University, Cleveland, OH 44106-7079, USA\\
$^3$ Fakult\"at f\"ur Physik, Universit\"at Bielefeld, Postfach 100131, D-33501 Bielefeld, Germany
}
\date{\today}

\begin{abstract}
  On  large  angular scales  ($\gtrsim$  60$^{\circ}$), the  two-point
  angular correlation  function of  the  temperature of  the cosmic  microwave
  background (CMB), as  measured (outside of the plane  of the Galaxy)
  by   the   {\it  Wilkinson   Microwave   Anisotropy  Probe},   shows
  significantly lower large-angle  correlations than expected from the
  standard inflationary cosmological  model. Furthermore, when derived
  from  the full CMB  sky, the  two lowest  cosmologically interesting
  multipoles, the  quadrupole ($\ell=2$) and  the octopole ($\ell=3$),
  are unexpectedly  aligned with each other.  Using randomly generated
  full-sky and  cut-sky maps,  we investigate whether  these anomalies
  are correlated at a statistically significant level. We conclusively
  demonstrate  that,   assuming  Gaussian  random   and  statistically
  isotropic  CMB anisotropies, there  is no  statistically significant
  correlation between the missing power on large angular scales in the
  CMB and the alignment of  the $\ell=2$ and $\ell=3$ multipoles.  The
  chance to  measure the sky  with both such a  lack of large-angle correlation  and 
such an alignment  of the  low multipoles is  thus quantified  to be
  below $10^{-6}$.
\end{abstract}

\maketitle

\section{Introduction}

Several anomalies in the cosmic microwave background maps observed by the WMAP
satellite \citep{Bennett2003,Jarosik2010} have been recently observed and much
discussed in the literature.  These include alignments of the lowest modes of
CMB anisotropy with each other, and with geometry and direction of motion of
the Solar System, as well as unusually low power at these largest
scales. Attempts to explain these anomalies include astrophysical,
instrumental, cosmological causes, as well as arguments that faulty data
analysis or {\it a posteriori} statistics are at work (for a review, see
\citet{CHSS_AA_review}). Two particularly puzzling features at large angular
scales are the alignments of large-angle anisotropies
\citep{oliveira04,Schwarz2004,Land2005a,lowl2}, and the low power at large
scales \citep{Spergel2003,wmap123}; each of these anomalies are $\ll 1\%$ likely.  
For reviews with differing points of view see \citet{wmap7_anomalies,CHSS_AA_review}.

A natural question that arises is whether the alignments and low power at
large scales are correlated. Naively, the answer is negative, since the
alignments are defined by orientation of the multipoles, and are independent
of the total power. In the language of multipole vectors, the normalization at
each multipole defines power, and is independent of the multipole vectors that
define orientations of multipoles. Such a conclusion has been verified, for
full sky maps, by \citet{Rakic2007}.

However, the answer for the {\it cut-sky} maps is less clear {\it a
  priori}. Consider the case where the large-scale multipoles are largely
planar (as is the case for our own quadrupole and octopole), thus creating
alignments. If the sky cut happens to be parallel to this plane, then the
cut-sky power may be unusually low, and this may well be happening more often
than when the cut is applied to unaligned (statistically isotropic)
maps. Clearly, the only safe way to investigate the correlation between the
low power and alignments in the presence of the realistic sky cut is to perform
Monte Carlo comparisons with Gaussian random, statistically isotropic
maps.

In this {\it Letter} we investigate the correlation between the alignments
and low power for the cut-sky as well as full-sky maps assuming a Gaussian
random, statistically isotropic, cosmological model. 

\section{Large-scale anomalies and WMAP7}\label{sec:anomalies}

\subsection{Low Power on Large Angular Scales}


The temperature anisotropy in the $\hat{\mathbf{e}} \left[=
  (\theta,\phi)\right]$ direction of the microwave sky can be represented by a
real scalar function ($\Delta T(\theta,\phi)$) on a sphere, expanded in terms
of the multipole moments
\vspace{-2mm}
\begin{equation}
\Delta T(\hat{\mathbf{e}}) \equiv \Delta T(\theta,\phi) = \sum_{\ell=0}^{\infty} T_\ell (\theta,\phi),
\end{equation}
where the $\ell$th multipole, $T_\ell (\theta,\phi)$, is expressed in
terms of the spherical harmonics, $Y_{\ell m}(\theta,\phi) $, as
\vspace{-2mm}
\begin{equation}
T_\ell (\theta,\phi)= \sum_{m=-\ell}^{\ell}a_{\ell m} Y_{\ell m}(\theta,\phi). 
\end{equation}

The standard inflationary cosmological model predicts that the fluctuations in
the microwave sky can be thought of as being sampled from a statistically
isotropic, Gaussian random field of zero mean. Gaussianity dictates that the
variances of these $a_{\ell m}$ would fully characterize the distribution and
statistical isotropy implies that these variances would depend only on $\ell$,
allowing us to write the {\it ensemble} average of any pair of $a_{\ell m}$ as
$\langle a_{\ell m}a^{*}_{\lpr \mpr}\rangle = C_{\ell} \delta_{\ell
  \lpr} \delta_{m \mpr}$, where $C_\ell$ is the (ensemble average
of) power in the $\ell$th multipole. 
In practice,
it is not possible to measure such an ensemble average as we have only one
realization of the universe. Instead, one uses its observable estimator,
$\hat{C}_\ell $, defined as
\vspace{-2mm}
\begin{equation}
\hat{C}_\ell \equiv \frac{1}{2\ell+1}\sum_{m=-\ell}^{\ell}|a_{\ell m}|^2.
\end{equation}
In the case of statistical isotropy, this $\hat{C}_\ell$, known as the angular power
spectrum, is an unbiased estimator of $C_\ell$. Additionally, if one
assumes Gaussianity, $\hat{C}_\ell $ is the best estimator of ${C}_\ell$ with
cosmic variance: ${\rm Var}(\hat{C}_\ell) = 2\hat{C}_\ell^2/(2\ell+1)$.

Instead of calculating the power in each multipole, one can consider the (real
space) two-point angular correlation function of the CMB temperature
fluctuations. Assuming statistical isotropy, one can express the angular
correlation function as $\mathcal{C}(\hat{\mathbf e}_1, \hat{\mathbf e}_2) =
\mathcal{C}(\theta) \equiv \langle T (\hat{\mathbf e}_1) T (\hat{\mathbf
  e}_2)\rangle_{\theta}$, where $\langle \cdot \rangle_{\theta}$ represents an
ensemble average over the temperatures $T (\hat{\mathbf e}_1) $ and
$T (\hat{\mathbf e}_2)$ in all pairs of directions $\hat{\mathbf e}_1$ and
$\hat{\mathbf e}_2$ separated by the angle $\theta$. One can get an unbiased
estimator of $\mathcal{C}(\theta)$ by replacing the ensemble average with the sky
average 
\begin{equation}
\label{cofthetaskyavg}
\hat{\mathcal{C}}(\theta) \equiv  \overline{T (\hat{\mathbf e}_1) T (\hat{\mathbf e}_2)}_{\theta}.
\end{equation}
Note that $\hat{C}_{\ell}$ and $\hat{\mathcal{C}}(\theta)$ only contain precisely
  the same information for full-sky data, and their analogues in the ensemble
  are informationally equivalent only if the sky is statistically isotropic.
Thus, simultaneous measurements of both $\hat{C}_\ell$
and $\hat{\mathcal{C}}(\theta)$ can be used to probe the validity of the assumption of
statistical isotropy.

\begin{figure}[!t]
\begin{center}
\includegraphics[scale=0.48]{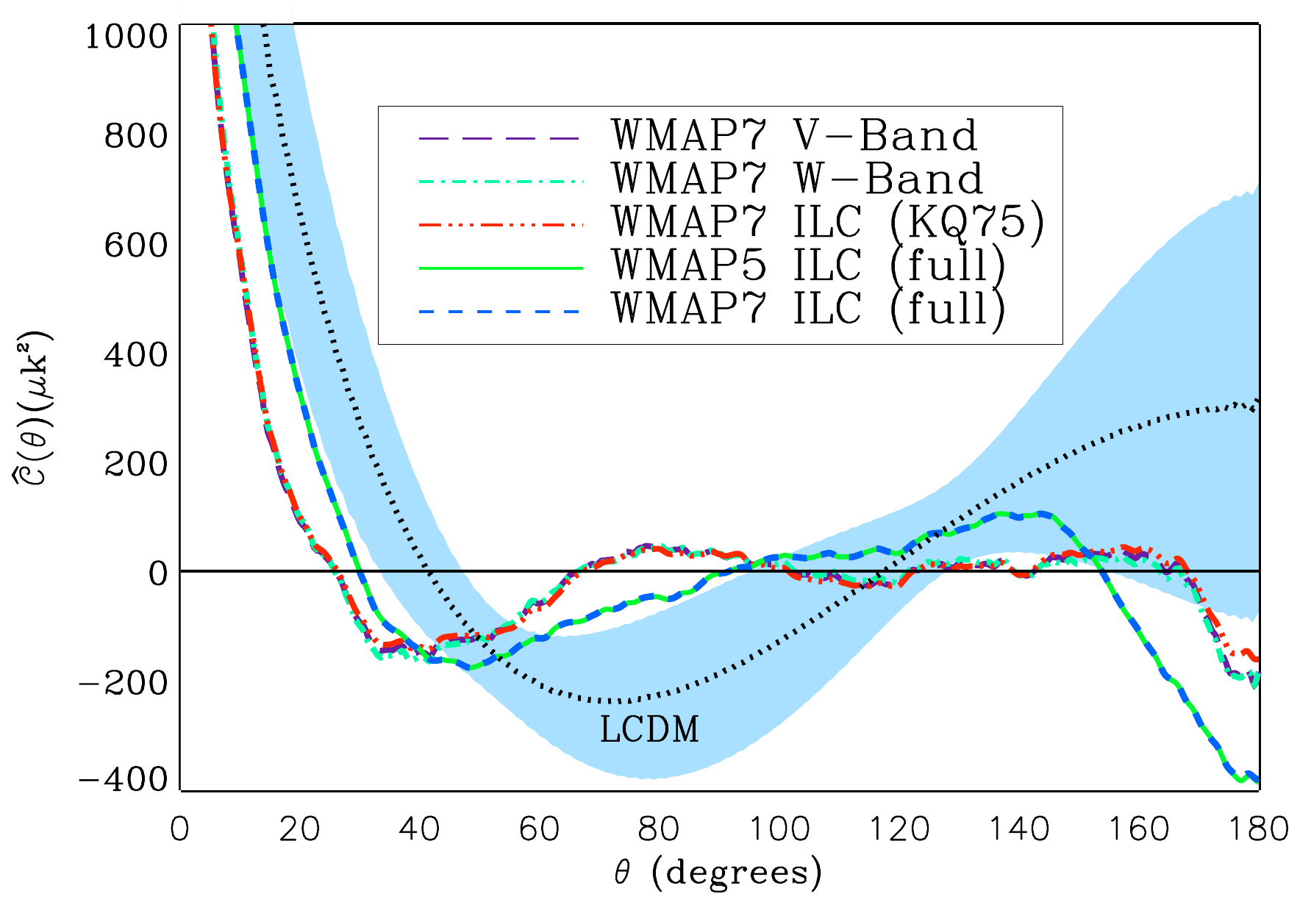}
\caption{The two point angular correlation function from the WMAP 7-year
  coadded data. The long-dashed and the dot-dashed lines show $\hat{\mathcal{C}}(\theta) \equiv
  \overline{T (\hat{\mathbf e}_1) T (\hat{\mathbf e}_2)}_{\theta}$ for the V
  and W bands, respectively, both masked with the KQ75y7 mask. The dashed line
  shows the correlation function for the ILC7 map (which covers the full sky). 
  The solid line shows the same for ILC5 for comparison.
  The $\hat{\mathcal{C}}(\theta)$ for the cut-sky version of the ILC7 map (using the same
  mask) is shown by the dot-dot-dot-dashed line. Finally, we show
  the angular two-point correlation function for the best-fit $\Lambda$CDM
  model with the dotted line, the 1$\sigma$ cosmic variance band around the
  best-fit being shown in blue.}
\label{S12}
\end{center}
\end{figure}

The two-point angular correlation function, $\hat{\mathcal{C}}(\theta)$, can be directly
measured as there is a very large number of independently measured pixels on
the WMAP sky. In Fig.~\ref{S12} we show $\hat{\mathcal{C}}(\theta) \equiv \overline{T
  (\hat{\mathbf e}_1) T (\hat{\mathbf e}_2)}_{\theta}$ (computed in pixel
space, using SpICE \citep{SpICE} at NSIDE=512) for the WMAP 7-year coadded
data. The results are shown for four different maps of WMAP7: the V and W
bands masked with the 7-year KQ75 mask (KQ75y7; henceforth),  the ILC map (which covers the full sky), and
the cut-sky version of the ILC map (using the same mask). For comparison, we
also show the full-sky ILC from the 5-year data.  Finally, we show
the angular two-point correlation function for the best-fit
$\Lambda$CDM model for WMAP7 data, along with the 1$\sigma$ cosmic variance band (in blue)
around the best-fit.  The most striking feature
of $\hat{\mathcal{C}}(\theta)$ for all the cut-sky maps is that they are very close to zero
for $\theta \gtrsim 60^{\circ}$, except for some anti-correlation near
180$^{\circ}$.

To quantify this lack of correlation, we follow past work
\citep{Spergel2003,wmap123,wmap12345} and adopt the $S_{1/2}$ 
statistic to test the total amount of correlation at angles
above $60^{\circ}$:
\vspace{-3mm}
\begin{equation}	
S_{1/2} \equiv \int_{-1}^{1/2} \left[ \hat{\mathcal{C}}(\theta)\right]^2 \mathrm{d}(\cos \theta).
\label{S12theta}
\end{equation}

The calculation of $S_{1/2}$ using Eq.~(\ref{S12theta}) is susceptible to
small-scale fluctuations in $\hat{\mathcal{C}}(\theta)$.  
To avoid this we evaluate $S_{1/2}$ via \citep{wmap12345} 
\begin{equation}
S_{1/2} = \frac{1}{(4\pi)^2}\sum_{\ell,\ell^{\prime}}(2\ell+1)(2\ell^{\prime}+1)
\hat{\mathcal{C}}_{\ell} {I}_{\ell,\ell^{\prime}} (1/2) \hat{\mathcal{C}}_{\ell^{\prime}},
\label{S12Cl}
\end{equation}
where $I_{\ell,\ell^{\prime}}(x) = \int_{-1}^x
P_\ell(x^{\prime})P_{\ell^{\prime}}(x^{\prime})dx^{\prime}$ and 
$\hat{\mathcal{C}}_{\ell}$ is the Legendre transform of 
$\hat{\mathcal{C}}(\theta)$
\vspace{-2mm}
\begin{equation}
  \hat{\mathcal{C}}_\ell \equiv 2\pi \int_{-1}^{1} P_\ell(\cos \theta)
  \hat{\mathcal{C}}(\theta) \mathrm{d} (\cos \theta).
\end{equation}
The $S_{1/2}$
values, calculated using Eq.~(\ref{S12Cl}), for different bands of WMAP 7-year
(as well as 5-year, for comparison) coadded data is shown in
Table~\ref{S12values}. Here we make use of SpICE at NSIDE=64.

\subsection{The Quadrupole-Octopole Alignment}

To measure the level of alignments between the quadrupole and octopole, we use
two of the most commonly used statistics: one based on the multipole vectors,
and one based on the angular momentum operator. As expected, the two
statistics are mutually highly correlated (see Table \ref{results} below); we
include both for completeness.


\begin{table*}[!t]
\caption{$S_{1/2}$-statistic from WMAP 5-year and WMAP 7-year Data}
\vspace{-3ex}
\begin{center}
\begin{tabular*}{0.98\textwidth}{@ {\extracolsep {\fill} }  c  |  c  c  c c c c c c c c}
\hline \hline
Data & V5 & W5 & ILC5 & \multirow{2}{*}{ILC5} & V7 & W7 & ILC7 & \multirow{2}{*}{ILC7} & \multirow{2}{*}{Theory5 $C_{\ell}$} & \multirow{2}{*}{Theory7 $C_{\ell}$} \\
Source & (KQ75y5)  & (KQ75y5) & (KQ75y5) &  & (KQ75y7) & (KQ75y7)  & (KQ75y7) & & & \\
\hline
S$_{1/2}$   & \multirow{2}{*}{1242}  & \multirow{2}{*}{1236} & \multirow{2}{*}{1054} & \multirow{2}{*}{8600} & \multirow{2}{*}{1312} & \multirow{2}{*}{1231} & \multirow{2}{*}{1141} & \multirow{2}{*}{8484} & \multirow{2}{*}{49040} & \multirow{2}{*}{46610}\\
($\mu$K$^4$) &  &  &  &  &  &  &  &  &  & \\
 \hline 
 \end{tabular*}
\label{S12values}
\end{center}
{\bf Note:} The values of the $S_{1/2}$-statistic are calculated for different
bands of WMAP 5-year and WMAP 7-year coadded data. We have used SpICE
\citep{SpICE} at NSIDE=64 and calculated $S_{1/2}$ using Eq.~(\ref{S12Cl}) with
$\ell_{max}=30$. The values of $S_{1/2}$ calculated from the best-fit theory
$C_{\ell}$ are also included for reference.
\end{table*}

Instead of expanding the temperature anisotropy of the microwave sky in
spherical harmonics, one can uniquely describe the temperature anisotropy in
the $\hat{\mathbf{e}} \left[= (\theta,\phi)\right]$ direction as \citep{Copi2004}
\vspace{-2mm}
\begin{equation}
T_\ell(\theta, \phi) =
A^{(\ell)} \left[ \prod_{i=1}^{\ell} 
\left(\hat{\mathbf{v}}^{(\ell,i)}\cdot \hat{\mathbf{e}} \right)-\mathcal{T}_\ell \right],
\label{mvformalism}
\end{equation}
where $A^{(\ell)}$ is a scalar which depends only on the total power in the
$\ell^{th}$ multipole and $\{\hat{\mathbf v}^{(\ell,i)}|i=1,...,\ell\}$ are $\ell$
`headless' unit vectors, known as the ``multipole vectors''. Here
$\mathcal{T}_\ell$ is the sum of all possible traces of the first term in the
right hand side of the Eq.~(\ref{mvformalism}).

To test the planarity of the quadrupole and octopole, for each multipole
$\ell$ we form the $\ell(\ell-1)/2$ cross-products (the ``oriented area''
vectors \citep{Schwarz2004})
\begin{equation}
{\mathbf w}^{(\ell;i,j)} = \hat{\mathbf v}^{(\ell,i)} \times \hat{\mathbf v}^{(\ell,j)},
\end{equation}
where the overall signs of the area vectors, ${\mathbf w}^{(\ell;i,j)}$, are
unimportant. For the following discussion, let us consider the lone area
vector for the quadrupole, ${\mathbf w}^{(2;1,2)}$, and the three area vectors
for the octopole: ${\mathbf w}^{(3;1,2)}$, ${\mathbf w}^{(3;2,3)}$, and
${\mathbf w}^{(3;3,1)}$.


To investigate the relative orientation of the quadrupolar plane with the
three octopolar planes, we can evaluate the magnitudes of the dot products
between ${\mathbf w}^{(2;1,2)}$ and ${\mathbf w}^{(3;i,j)}$, given by
\vspace{-2mm}
\begin{equation}
A_{1 \; \{ \text{or} \; 2 \; \text{or} \; 3\}} \equiv \left| {\mathbf w}^{(2;1,2)}\cdot {\mathbf w}^{(3;1,2) \; \{\text{or} \;(3;2,3) \; \text{or} \; (3;3,1)\}} \right|.
\end{equation}
High values of $A_i$
imply that the $\ell=2$ and $\ell=3$ planes are aligned with each other. One can
test the combined planarity of the quadrupole and octopole by using the
statistic that takes the average of the dot products
\vspace{-2mm}
\begin{equation}
S^{(3,q)} \equiv \frac{1}{q}\sum_{i=1}^{q} A_i.
\end{equation}
Using ILC7 and ILC5, we find $S^{(3,3)}=0.736$ and $S^{(3,3)}=0.753$, respectively. 
In the subsequent sections, to be conservative, we use $S^{(3,3)}=0.798$ \citep{wmap123}, obtained  
using the cleaned full sky map of \citet{TOH} (TOH; henceforth).






An alternative statistic that tests the planarity is given by the angular
momentum dispersion (\cite{oliveira04}).  For each $\ell$, an axis $\hat{\mathbf
  n}_\ell$ can be found, around which the angular momentum dispersion
\vspace{-2mm}
\begin{equation}
  \left\langle T_{\ell}(\hat{\mathbf n}_\ell) | 
(\hat{\mathbf n}_\ell \cdot \mathbf{L})^2 | T_{\ell}(\hat{\mathbf
  n}_\ell) 
\right\rangle = \sum_{m=-\ell}^{\ell} m^2 |a_{\ell m}(\hat{\mathbf n}_\ell)|^2
\end{equation}
is maximized. Here $\mathbf{L}$ is the angular momentum operator and $a_{\ell
  m}(\hat{\mathbf n}_\ell)$ are the spherical harmonic coefficients of the CMB
map in coordinate system with its $z$-axis in the $\hat{\mathbf n}_\ell$
direction. Note that this angular momentum axis distills a limited amount of
information from each multipole, reducing the d.o.f.\ to just 2 from the
$2\ell$ d.o.f. in $\hat{\mathbf v}^{(\ell,i)}$.

\citet{lowl2} found that $\hat{\mathbf n}_2 \cdot \hat{\mathbf n}_3=
0.962$ for the TOH map. 
The corresponding values obtained from the analysis of the
WMAP 7-year data is 0.937.  The measured values of $S^{(3,3)}$ and
$\hat{\mathbf n}_2 \cdot \hat{\mathbf n}_3$ consistently indicate that the
quadrupole and the octopole are unexpectedly aligned, both in the 5 and 7 year
WMAP data.

\section{In Search for a Correlation} \label{sec:search}

We have just seen that WMAP7 confirms results from its earlier data releases:
at large angular scales ($\gtrsim$ 60$^{\circ}$), the two-point correlation
function of the temperature of the CMB, as measured by WMAP, is significantly
smaller in magnitude than expected from the predictions of the standard inflationary
cosmological model; and in addition, the planes of the quadrupole ($\ell=2$) and the
octopole ($\ell=3$) are unexpectedly aligned with each other. In this section, we
investigate whether these anomalies are correlated at a statistically
significant level.

More specifically, we want to investigate a couple of different scenarios:
\begin{itemize}
\item If the Gaussian random, statistically isotropic CMB maps are constrained
  to have low angular correlation on large angular scales, are they more likely to exhibit planarity
  and alignment of the quadrupole and octopole?
\item If the Gaussian random, statistically isotropic CMB maps are constrained
  to have aligned quadrupole and octopole, are they more likely to have low
  angular correlation on large angular scales?
\end{itemize}

Although the answers to the above questions might seem to be obvious for the
full-sky case, there is no direct reason why no such correlation should exist
for the case of cut-sky maps. In what follows, we attempt to address each of
these questions for both the full-sky and the cut-sky cases.

\subsection{Monte-Carlo Analysis}

We first generate 100,000 constrained (low angular correlation on large angular scales) realizations
of the CMB sky with selection criterion of $S_{1/2} < 8583 (\mu K)^4$
(i.e.\ lower than the ILC5 full-sky $S_{1/2}$ from \cite{wmap12345}) {\it and} the
cut-sky $S_{1/2} < 1152 (\mu K)^4$ (i.e.\ lower than the ILC5 cut-sky
map).\footnote{It should be noted that these values of $S_{1/2}$ are
  corresponding to the WMAP 5-year ILC full sky (DQ corrected) and WMAP 5-year
  ILC cut sky (using KQ75y5 mask, DQ corrected) maps, since we started this
  analysis prior to WMAP 7-year data release. In fact, these values are
  comparable to the corresponding values for the WMAP 7-year results shown in
  Table~\ref{S12values}.} This is our sample of Gaussian random, statistically
isotropic maps that are constrained to have low angular correlation on large scales, and we use them to test
the probability of alignments under this constraint, i.e., the probability of
having $S^{(3,3)} > 0.798$ for these low-correlation maps.  We also calculate the
probability of these low--correlation maps to have aligned angular momentum
dispersion axes, i.e., $\hat{\mathbf n}_2 \cdot \hat{\mathbf n}_3 >
0.962$.  We then compare these probabilities with the corresponding ones for
the unconstrained maps.

\begin{table}[t!]
\begin{center}
 \caption{Probabilities of having Low-Power on Large Angular Scales and Quadrupole-Octopole Alignment for the Unconstrained and Constrained Maps }
\vspace{-2.5ex}
\begin{tabular*}{0.48\textwidth}{@ {\extracolsep {\fill} }  c || c | c c}
\hline\hline
Probabilities & Unconstrained & \multicolumn{2}{c}{Constrained Maps} \\
    (in $\%$) &    Maps       & Low Power      & Full-Sky Aligned\\
\hline \hline
$P\left(S_{1/2}^{{\rm full\, sky}}\right)$ & 7.0  &   ---     & {\bf 6.9} \\
$P\left(S_{1/2}^{\rm cut\, sky}\right)$    & 0.05 &   ---     & {\bf 0.07}\\
$P\left(S^{(3,3)}\right)$                & 0.12 & {\bf 0.12} & ---\\
$P\left(\hat{\mathbf n}_2 \cdot \hat{\mathbf n}_3\right)$ 
                                        & 0.37 & {\bf 0.36} & {\bf 99.6}  \\
\hline\hline
 \end{tabular*} 
 \label{results}
 \end{center}
{\bf Note:} The statistics $S^{(3,3)}$ and $\hat{\mathbf n}_2
   \cdot \hat{\mathbf n}_3$ are always evaluated on the full sky. The cases with the dashes
   denote circular comparisons.
 \end{table}

To test the probability of low angular correlation on Gaussian random, statistically
isotropic maps that are constrained to be aligned, we generate 100,000 MC maps
constrained to have $S^{(3,3)} > 0.798$, calculated using the full sky as the
cut sky alignments are extremely difficult to test (the errors in multipole
vectors become large for a cut larger than a few degrees). We then evaluate
the $S_{1/2}$ statistic for these constrained maps 
(with and without applying
the WMAP 5-year KQ75 mask (KQ75y5)) and calculate the percentage of these maps having $S_{1/2} <
8583 (\mu K)^4$ (for the full-sky case) or $S_{1/2} < 1152 (\mu K)^4$ (for the
case using KQ75y5).  We also calculate the probability of these constrained
maps to have $\hat{\mathbf n}_2 \cdot \hat{\mathbf n}_3 > 0.962$. As
discussed before, we then compare these probabilities with the corresponding
ones for the unconstrained maps.

\subsection{Results and Discussions}

Our main results are summarized in Table~\ref{results}. Using our sample of
100,000 low-power maps, we find the probability of these maps to have
$S^{(3,3)} > 0.798$ is $0.12\%$, which is the same as the probability of
completely unconstrained maps to have quadrupole-octopole alignment
\citep{lowl2}.  This shows that the low-power maps do not have a higher
probability to have an alignment of the $\ell=2$ and $\ell=3$ planes. The
probability of these low-power maps to have aligned angular momentum
dispersion axes ($\hat{\mathbf n}_2 \cdot \hat{\mathbf n}_3 > 0.962$) is
also very low ($0.36\%$) and essentially identical as that found for unconstrained maps
(0.37\%; \citet{lowl2}).

Using 100,000 maps constrained to be have quadrupole-octopole alignment
($S^{(3,3)} > 0.798$), we find $6.9\%$ probability for these maps to have
$S_{1/2} < 8583 (\mu K)^4$ for the full-sky case. Comparing this with the
probability ($7.0\%$) of having low angular correlation for completely unconstrained maps,
we find no correlation between alignment and low angular correlation for the full-sky case.
 
For the cut-sky case (using the KQ75 mask), however, we find $0.07\%$
probability for these maps to have $S_{1/2} < 1152 (\mu K)^4$.  This
probability is slightly higher ($\gtrsim 2\sigma$ with $\sigma$ a Poisson
error) than $P\left(S_{1/2}^{\rm cut\,
    sky}\right) = 0.05\%$ for unconstrained maps. However, this difference is
not statistically significant to conclude any detectable correlation between
alignment and low power for cut skies.

\section{Conclusion}
\label{sec:conclude}

Applying the multipole vector formalism to Gaussian random and statistically
isotropic realizations of CMB maps (both full-sky and cut-sky) lacking
correlation on large angular scales, we find no increased probability for
alignment of the quadrupole and the octopole than expected in the case of
unconstrained random maps. These low-power maps also do not show higher
angular momentum dispersion. On the flip side, we also find that realizations
of CMB maps (both full-sky and cut-sky), constrained to have aligned
quadrupole and octopole at a level greater than that exhibited by WMAP, do not
have lower power on large scales than expected in the case of unconstrained
random realizations.

Our results conclusively demonstrate that, under the standard Gaussian and
isotropic model, there is no statistically significant correlation between the
missing power on large angular scales in the CMB and the alignment of the
$\ell=2$ and $\ell=3$ multipoles. Therefore, in the context of the standard
model, their combined statistical significance is equal to the product of
their individual significances. For example, simultaneous observation of the
missing large-angle correlations with probability $P(S_{1/2}^{\rm cut\,
  sky})\lesssim 0.1\%$ {\it and} alignments with the probability
$P(S^{(3,3)})\simeq 0.1\%$ is likely at the $\lesssim 0.0001\%$ level.  Given
that both anomalies occur at the largest observable scales and are correlated
with special directions in the sky (ecliptic and/or dipole), they clearly
require a causal explanation.

\section*{Acknowledgments} 
We are supported by the NSF under contract AST-0807564 (DS and DH), NASA under
contracts NNX09AC89G (DS, DH and CJC) and NNX07AG89G (CJC and GDS), DOE OJI
grant under contract DE-FG02-95ER40899 (DH), DOE theory grant to CWRU (GDS),
and by a DFG grant (DJS).
\vspace{0.5cm}

\bibliography{cmb_review}

\end{document}